\documentstyle[multicol,aps]{revtex}
\input epsf.tex
\begin{document}
\draft
\title{Formation of heavy quasiparticle state in two-band Hubbard model}
\author{H. Kusunose\cite{kusu}}
\address{Department of Physics, Tohoku University, Sendai, 980-8578, Japan}
\author{S. Yotsuhashi and K. Miyake}
\address{Department of Physical Science, Osaka University, Toyonaka, 560-8531, Japan}
\date{October 4, 1999}
\maketitle
\begin{abstract}
A realization of heavy fermion state is investigated on the basis of two-band Hubbard model.
By means of the slave-boson mean-field approximation, it is shown that for the intermediate electron density, $n_e=1.5$, the inter-band Coulomb repulsion $U$ strongly emphasizes initially small difference between bands, and easily stabilizes integral valence in the lower band.
As a result, a strong renormalization takes place in the lower band and the mixing strength between two bands.
It gives rise to a sharp peak at the Fermi level in the quasiparticle density of states, as that obtained in the periodic Anderson model.
In contrast to a simple insight that the Hund's-rule coupling $J$ reduces the characteristic energy, it  turns out to be almost irrelevant to the renormalization for $J<U$.
The required conditions are suitable for LiV$_2$O$_4$, the first observed heavy fermion compound in transition metal oxide.
\end{abstract}

\begin{multicols}{2}
%
%
\section{Introduction}
The recent discovery of heavy fermion behavior in LiV$_2$O$_4$\cite{kondo97} uncovers the latent possibilities to explore Kondo physics in the lattice $d$-electron systems, which is restricted so far to $f$-electron systems containing lanthanide or actinide atoms\cite{grewe91,hewson93}.
The heavy fermion behavior has been widely observed in various measurements, such as specific heat\cite{johnston99}, susceptibility\cite{kondo99}, $^7$Li and $^{51}$V NMR\cite{mahajan98,fujiwara98}, $\mu$SR\cite{kondo97,merrin}, thermal expansion\cite{chmaissem97}, quasielastic neutron scattering\cite{krimmel99}, resistivity\cite{takagi99} and so on\cite{johnston99a,kondo00}.
The low-energy physics is characterized by the large mass enhancement in the specific heat coefficient, $\gamma\sim 0.21$ J/V-mol K$^2$ with the Sommerfeld-Wilson ratio $R_{\rm W}\sim 1.71$.
The characteristic temperature of Kondo or spin fluctuation is estimated as $T^*\sim 30$ K.
With elevated temperature, the magnetic susceptibility approximately follows the Curie-Weiss law, $C/(T-\theta)$, where the Curie constant $C$ is consistent with a $V^{+4}$ spin $S=1/2$ with $g$ factor $2.23$, and the negative Weiss temperature ($\theta=-63$ K) is familiar to $f$-electron heavy fermions.

The several band structure calculations have been made at present\cite{eyert99,anisimov99,matsuno99,singh99}.
They have revealed that the octahedral coordination of the oxygen ions around the V atom causes the large splitting of $d$ states into $t_{2g}$ and $e_g$ orbitals.
The partially filled $t_{2g}$ bands can be described roughly by V-V hopping, and they are well separated by the filled O-$2p$-bands and the empty $e_g$ bands.
Eyert et al. suggest the specific heat enhancement comes from spin fluctuations with the magnetic order suppressed by the geometric frustration\cite{eyert99}.
In similar context, spin fluctuation nearby magnetically unstable point in Li$_{1-x}$Zn$_x$V$_2$O$_4$ is discussed by Fujiwara et al.\cite{fujiwara98,fujiwara99,ueda97}.

On the other hand, an attempt to map onto the conventional periodic Anderson model (PAM) is made by Anisimov et al.\cite{johnston99,kondo99,mahajan98,anisimov99}.
In the realistic treatment of the trigonal symmetry crystal field, triply degenerate $t_{2g}$ orbitals split into the non-degenerate $A_{1g}$ and doubly degenerate $E_g$ representations of the $D_{3d}$ group.
Their assertion is that due to the Coulomb interaction among $d$ electrons, one electron of the $d^{1.5}$ configuration is localized into the $A_{1g}$ orbital and the rest partially fills a relatively broad conduction band made from $E_g$ orbitals.

The idea seems to resolve the enormous differences from the isostructural LiTi$_2$O$_4$, which has 0.5 $d$ electrons per Ti atom.
It shows the relatively $T$-independent Pauli paramagnetism in susceptibility and the superconducting state below $T_c=13.7$ K, which is well described by the BCS theory\cite{johnston76,mccallum76}.
The related discussion is also made by Varma\cite{varma99}.

Nevertheless, it is not easy to for $d$ electron be localized because of the much larger spatial extent of $d$ orbitals than of $f$ orbitals, unless the $A_{1g}$ orbital is located much deeper than the $E_g$ band.
In other words, the use of the PAM has no solid ground contrary to a naive expectation.
Moreover, the intra-band Coulomb repulsion does not work effectively to make heavy mass since each electron favors to place in different orbitals rather than in the same ones.

In this paper, we clarify how a heavily renormalized quasiparticle ground state is realized in the two-band Hubbard model, leading to somewhat different physical picture from that argued in Ref. \cite{anisimov99}.
By using the slave-boson mean-field approximation\cite{kotliar86,trees95}, we discuss the importance of the inter-band Coulomb repulsion to make one of the $d$ electrons be technically localized and provide a situation similar to that given by the PAM.
Since the geometric frustration is expected to prevent a long-range order, we restrict our attention to a paramagnetic ground state.
In the same stand, we argue about LiTi$_2$O$_4$ as the case of smaller electron density, $n_e=0.5$.

%
%
\section{Model and Formulation}
Let us start with two-band Hubbard model, which may capture the broad $E_g$ band (called as $A$) with the width 2 eV and the narrow $A_{1g}$ band (called as $B$) with the width 1 eV.
The latter is located at 0.1 eV lower than the former due to the trigonal distortion.
The $d$-$d$ Coulomb interactions and the Hund's-rule coupling are estimated as 3 eV and 1 eV, respectively\cite{anisimov99}.
The mixing strength between bands must be much smaller than the width of both bands.
The Hamiltonian is given by
\begin{eqnarray}
&&H=\sum_{k\ell\sigma}\left[\left(\epsilon_{k\ell}+E_\ell-\mu\right)c^\dagger_{k\ell\sigma}c_{k\ell\sigma}+vc^\dagger_{k\ell\sigma}c_{k\bar{\ell}\sigma}\right]+H_{\rm int},
\nonumber\\&&\mbox{\hspace{5mm}}
H_{\rm int}=\sum_{i\ell}U_\ell n_{i\ell\uparrow}n_{i\ell\downarrow}+U\sum_{i\alpha\beta}c^\dagger_{iA\alpha}c_{iA\alpha}c^\dagger_{iB\beta}c_{iB\beta}
\nonumber\\&&\mbox{\hspace{1.5cm}}
-\frac{J}{4}\sum_{i\alpha\beta\gamma\delta}c^\dagger_{iA\alpha}c_{iA\beta}c^\dagger_{iB\gamma}c_{iB\delta}\vec{\sigma}_{\alpha\beta}\cdot\vec{\sigma}_{\gamma\delta}.
\label{hint}
\end{eqnarray}
The first term denotes the kinetic energy of conduction electrons for bands $\ell$, in which $E_\ell=\pm\Delta/2$ for $A$ and $B$ bands, $\Delta$ being the trigonal splitting, and $\mu$ the chemical potential.
The second is a mixing strength between bands and its {\it k}-dependence is neglected for simplicity.
The intra- and inter-band Coulomb interactions $U_\ell$ and $U$ as well as the Hund's-rule coupling $J$ at the site $i$ are considered in $H_{\rm int}$, where the intra-band Coulomb interactions are set by $U_A=U_B=\infty$ for simplicity.

To solve this Hamiltonian, we introduce slave-boson fields for $d^0$-$d^2$ states at each site.
We associate a boson $e_i$ for $d^0$ state, $p_{i\ell\sigma}$ for $d^1$ states, and $d_{iSS_z}$ for $d^2$ states labeled by their spin states, $(S,S_z)$, respectively.
For the uniform solution of the mean-field approximation, we replace these bosons by site-independent $c$ numbers.
Assuming the paramagnetic ground state, five bosons, $e=e_i$, $p_A=p_{iA\sigma}$, $p_B=p_{iB\sigma}$, $d_0=d_{00}$ and $d_1=d_{1S_z}$, are involved in calculation.

The completeness relation for $d^0$-$d^2$ states is given by
\begin{equation}
I-1=0,\;\;I\equiv e^2+2\sum_\ell p_{\ell}^2+d_0^2+3d_1^2.
\end{equation}
Since the probabilities for the singlet and the triplet states are given by $d_0^2$ and $3d_1^2$, respectively, the two-body interactions are rewritten in terms of bosons as
\begin{equation}
H_{\rm int}^{\rm MF}/N=U\left(d_0^2+3d_1^2\right)-\frac{3}{4}J\left(d_1^2-d_0^2\right),
\end{equation}
where $N$ is the number of sites.
The each species of electrons must satisfy the constraint at each site,
\begin{equation}
c^\dagger_{i\ell\sigma}c_{i\ell\sigma}-Q_{\ell}=0,\;\;Q_{\ell}\equiv p_\ell^2+\frac{1}{2}\left(d_0^2+3d_1^2\right).
\end{equation}

In this slave-boson scheme\cite{kotliar86}, the hopping term is suppressed as
\begin{equation}
c^\dagger_{k\ell\sigma}c_{k\ell\sigma}\to q_\ell c^\dagger_{k\ell\sigma}c_{k\ell\sigma},\;\;\;
c^\dagger_{kA\sigma}c_{kB\sigma}\to q c^\dagger_{kA\sigma}c_{kB\sigma},
\end{equation}
where the renormalization factors $q_\ell$ and $q$ are given by
\begin{eqnarray}
&&
q_\ell=\tilde{z}_\ell^2,\;\;\;q=\sqrt{q_A q_B},\;\;\;\tilde{z}_\ell=Q_\ell^{-1/2}z_\ell\left(1-Q_\ell\right)^{-1/2},
\\&&
z_\ell=e\; p_\ell+\frac{1}{2}\left(d_0+3d_1\right)\;p_{\bar{\ell}}.
\label{zl}
\end{eqnarray}
Note that the Gutzwiller correction $Q_\ell^{-1/2}(1-Q_\ell)^{-1/2}$ is necessary to reproduce non-interacting limit\cite{kotliar86,buenemann98}.

Finally, we obtain the mean-field free energy per site with two Lagrange multipliers $\lambda$ and $\lambda_\ell$,
\begin{eqnarray}
F^{\rm MF}/N&=&-\frac{2}{\beta N}\sum_{k}\sum_m^\pm \ln\left(1+e^{-\beta\left(\tilde{E}_{km}-\mu\right)}\right)+H_{\rm int}^{\rm MF}/N
\nonumber\\&&
+\lambda\left(I-1\right)-2\sum_\ell\lambda_\ell Q_\ell,
\end{eqnarray}
where the bonding and the anti-bonding ($m=\mp$) quasiparticle bands are given by
\begin{eqnarray}
&&
\tilde{E}_{km}=\frac{1}{2}\left[\tilde{\xi}_{kA}+\tilde{\xi}_{kB}+m\sqrt{(\tilde{\xi}_{kA}-\tilde{\xi}_{kB})^2+4\tilde{v}^2}\right],
\nonumber\\&&\mbox{\hspace{10mm}}
\tilde{\xi}_{k\ell}\equiv q_\ell\epsilon_{k\ell}+E_\ell+\lambda_\ell,\;\;\;\;\tilde{v}\equiv qv.
\label{qb}
\end{eqnarray}
The width of the band $\ell$ and the mixing strength are renormalized by factors $q_\ell$ and $q$, respectively.
The position of the bands are moved up by an amount $\lambda_\ell$.

Minimizing the free energy with respect to five bosons and two Lagrange multipliers, the set of self-consistent equations are obtained as follows:
\begin{eqnarray}
&&
e^2+2\sum_\ell p_\ell^2+d_0^2+3d_1^2-1=0,
\label{sceq1}
\\&&
\bar{n}_\ell-2[p_\ell^2+\frac{1}{2}(d_0^2+3d_1^2)]=0,
\label{sceq2}
\\&&
\sum_\ell \frac{\bar{\epsilon}_\ell}{2} \frac{\partial q_\ell}{\partial e}+v\bar{r}\frac{\partial q}{\partial e}+\lambda e=0,
\label{sceq3}
\\&&
\sum_\ell \frac{\bar{\epsilon}_\ell}{2} \frac{\partial q_\ell}{\partial p_{\ell'}}+v\bar{r}\frac{\partial q}{\partial p_{\ell'}}-2(\lambda_{\ell'}-\lambda)p_{\ell'}=0,
\label{sceq4}
\\&&
\sum_\ell \frac{\bar{\epsilon}_\ell}{2} \frac{\partial q_\ell}{\partial d_0}+v\bar{r}\frac{\partial q}{\partial d_0}+(T_0-\sum_\ell\lambda_\ell+\lambda)d_0=0,
\label{sceq5}
\\&&
\sum_\ell \frac{\bar{\epsilon}_\ell}{2} \frac{\partial q_\ell}{\partial d_1}+v\bar{r}\frac{\partial q}{\partial d_1}+3(T_1-\sum_\ell\lambda_\ell+\lambda)d_1=0,
\label{sceq6}
\\&&
n_e-\sum_\ell\bar{n}_\ell=0,
\label{sceq7}
\end{eqnarray}
where the energies of the singlet and the triplet states are defined by $T_0=U+3J/4$ and $T_1=U-J/4$.
The last equation is responsible for determining the chemical potential for given electron density $n_e$.
Hereafter we restrict ourselves to the case at zero temperature $\beta^{-1}=0$. At zero temperature the averages of electron densities, mixing amplitude and kinetic energies are given by
\begin{eqnarray}
&&
N\bar{n}_\ell\equiv\sum_{k\sigma}\left\langle c^\dagger_{k\ell\sigma}c_{k\ell\sigma}\right\rangle=\sum_{km\sigma}\rho^m_{k\ell}\theta(\mu-\tilde{E}_{km}),
\\&&
N\bar{r}\equiv\sum_{k\sigma}\left\langle c^\dagger_{kA\sigma}c_{kB\sigma}\right\rangle=\sum_{km\sigma}\zeta^m_{k}\theta(\mu-\tilde{E}_{km}),
\\&&
N\bar{\epsilon}_\ell\equiv\sum_{k\sigma}\epsilon_{k\ell}\left\langle c^\dagger_{k\ell\sigma}c_{k\ell\sigma}\right\rangle=\sum_{km\sigma}\epsilon_{k\ell}\rho^m_{k\ell}\theta(\mu-\tilde{E}_{km}),
\end{eqnarray}
with
\begin{eqnarray}
&&
\rho^m_{k\ell}=\frac{1}{2}\left[1+m\frac{\tilde{\xi}_{k\ell}-\tilde{\xi}_{k\bar{\ell}}}{\sqrt{(\tilde{\xi}_{kA}-\tilde{\xi}_{kB})^2+4\tilde{v}^2}}\right],
\\&&
\zeta^m_k=\frac{m\tilde{v}}{\sqrt{(\tilde{\xi}_{kA}-\tilde{\xi}_{kB})^2+4\tilde{v}^2}}.
\end{eqnarray}
For simplicity, we use a rectangular density of states (DOS) with a linear dispersion relation, i.e., $\epsilon_{k\ell}=W_\ell x/2$ for $|x|\le 1$. Then, The $k$-summation in the averages can be carried out analytically with the integration, $1/N\sum_{k}\to 1/2\int^1_{-1}dx$.

%
%
\section{Results and Discussions}
The set of self-consistent equations are solved numerically.
In the following we use parameters, $W_A=2$, $W_B=1$, and $v=0.2$ (eV).
Figure \ref{fig1} shows the quasiparticle DOS with and without interactions, $U_{\ell}$, $U$ and $J$, for $\Delta=0.2$ (eV) and $n_e=1.5$ corresponding to LiV$_2$O$_4$.
The bandwidth is renormalized slightly by the intra-band Coulomb repulsion $U_A=U_B=\infty$.
However, the renormalization amplitude of the narrower $B$ band, $q_B$, remains at the order of $10^{-1}$ without the inter-band repulsion, since each electron favors to place in different orbitals rather than in the same ones.
On the contrary, with the inter-band interactions $U=3$ and $J=1$ (eV), a strong renormalization takes place and it gives rise to a sharp peak at the Fermi level in the quasiparticle DOS. (Its width is about 40 K.)
Note that both the upper and the lower Hubbard bands cannot be argued in the mean-field approximation.
The inset in Fig. \ref{fig1} shows the quasiparticle DOS for the case of $n_e=0.5$ corresponding to LiTi$_2$O$_4$.
As it is expected, the renormalization is very weak, $q_B\sim 0.7$.

To elucidate why the inter-band interaction assists the strong renormalization, we discuss the limiting cases for $n_e=1.5$.
In the absence of $H_{\rm int}$ in eq. (\ref{hint}), a rather large trigonal splitting, i.e., $\Delta\sim W_A/4$ is required to stabilize the integral valence in the $B$ band, $\bar{n}_B$.
While, in the case of the strong repulsion, $U_A=U_B=\infty$, and $U/W_\ell\gg1$, the inter-band Coulomb repulsion considerably enhances the difference of electron densities between two bands, $\Delta n=\bar{n}_B-\bar{n}_A$, because of the relation $U\bar{n}_A\bar{n}_B=U[n_e^2-(\Delta n)^2]/4$.
In this case with $d_0\sim d_1$ for $J/U\ll 1$, the renormalization factor vanishes as
\begin{equation}
q_B\sim\frac{n_e-1}{\bar{n}_B}\frac{1-\bar{n}_B}{1-\bar{n}_B/2}.
\label{rf}
\end{equation}
Note that in the PAM with $U=\infty$ the hybridization between $f$ and conduction electrons is suppressed as $V^2\to q_f V^2$, where $q_f=(1-n_f)/(1-n_f/2)$\cite{kotliar86,rice85,newns87}.
We emphasize here that although the mechanism of strong renormalization appears similar to that in the PAM, it is totally a new mechanism that the Kondo limit is {\it dynamically} provided by the inter-band Coulomb repulsion.
It will be shown below that the integral $\bar{n}_B$ can be stabilized even for rather small $\Delta$, which is based on the detailed balance between the kinetic energy and the inter-band Coulomb interaction.

On the other hand, in the case of $J\gg (U,W_\ell)$, the amplitude of triplet state, $d_1$, becomes as large as possible.
Its maximum is bounded by $3d_1^2\le \min(\bar{n}_A,\bar{n}_B)$.
Thus, in order to take the largest value, $\Delta n$ is suppressed.
In the limit of large $J$, the probabilities for $d^0$-$d^2$ states, $(e^2,p_\ell^2,3d_1^2)$ approach $(1-n_e/2,0,n_e/2)$, respectively.
Namely, the system undergoes a dimerization with a charge order, and the renormalization factor $q_\ell$ vanishes since $p_\ell\to 0$ in eq. (\ref{zl}).

Figure \ref{fig2} shows $\bar{n}_B$ as a function of the trigonal splitting $\Delta$.
The intra-band Coulomb interaction $U_A$ and $U_B$ (circle) somewhat enhances $\bar{n}_B$.
To stabilize the integral valence, however, $\Delta$ is required as large as that for the case of free electrons (square), i.e., $\Delta\sim 0.5$ eV.
On the other hand, the inter-band interaction (triangle) works effectively to stabilize integral valence even for small $\Delta$.
Note that $\bar{n}_B$ is almost unity for $\Delta\sim 0.1$ eV.

The $\Delta$ dependence of the renormalization factors are shown in Fig. \ref{fig3}.
As expected from the above discussion, in the presence of the inter-band interactions (square), $B$ band is highly renormalized owing to $\bar{n}_B\to 1$, while the intra-band interactions are almost irrelevant up to relatively large $\Delta$ (circle).
It is noted that the upper band is almost unrenormalized, while the band mixing is also renormalized considerably, i.e., $q=\sqrt{q_Aq_B}$.

In order to elucidate how the inter-band interactions reduce renormalization factors, we extract the interaction dependences of the renormalization factors for $\Delta=0.2$ in Fig. \ref{fig4}.
It is shown that the inter-band Coulomb interaction (square) effectively reduces renormalization factors.
On the other hand, the Hund's-rule coupling turns out to be almost irrelevant for $J<U$ (circle, triangle).
This is in contrast to a simple insight that a strong cancellation between the Hund's-rule coupling and the Kondo exchange coupling considerably reduces the characteristic energy as discussed in Refs. \cite{okada73,yotsuhashi}.
Discussions are based on the impurity model and hence there is no constraint for the electron density such as $n_e=\bar{n}_A+\bar{n}_B$.
Since a change of Hund's-rule coupling requires another change of parameters to restore a given electron density, all parameters must be treated in a self-consistent fashion.
This might remove the discrepancy from our result that the Hund's-rule coupling is almost irrelevant in renormalization.

%
%
\section{Summary}
We have shown that the inter-band Coulomb repulsion plays a significant role to reduce the renormalization factor strongly, while the Hund's-rule coupling is almost irrelevant in renormalization for $J<U$.
Even though both bands are rather broad and a splitting between bands is very small, the resultant quasiparticle can have a heavy mass enhanced by about $10^2$ times since the Kondo limit is dynamically provided by the inter-band Coulomb repulsion.
Although the values obtained by slave-boson approach may be changed quantitatively by more elaborate one, the situation is highly plausible to account for the heavy-fermion behavior in LiV$_2$O$_4$ and the enormous differences from LiTi$_2$O$_4$.

Since the heavily renormalized quasiparticle has been stabilized dynamically by the inter-band Coulomb repulsion, it should couple strongly with orbital fluctuations at higher temperature.
The large contribution to the specific heat observed above $T^*$\cite{johnston99} is presumably related to the orbital fluctuations.

At low temperature $d$ electron systems generally exhibit a long-range order.
If a paramagnetic state survives due to some reasons such as a geometric frustration, one would expect heavy-fermion behavior in numerous $d$ electron systems. The resultant quasiparticle holds the possibility of showing fascinating phenomena such as a novel superconductivity mediated by orbital fluctuations.

%
%
\acknowledgments
H. K. would like to thank H. Yamagami for fruitful discussion on electric structure calculation. He has also benefited from conversations with D. L. Cox and S. Kondo.
This work was supported by a Grant-in-Aid for the encouragement of Young Scientists and also in part by a Grant-in-Aid for COE Research (10CE2004) from the Ministry of Education, Science, Sports and Culture of Japan.
K.M. acknowledges C. M. Varma for leading his attention to this problem.
%
%

\begin{minipage}{8.2cm}
\begin{figure}
\begin{center}
\epsfxsize=8cm \epsfbox{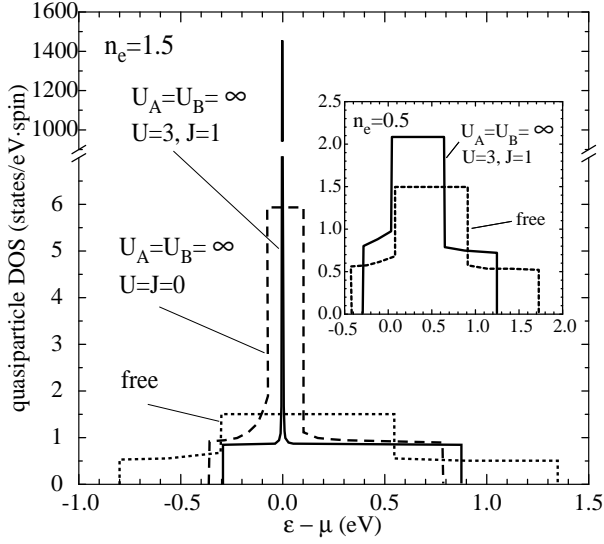}
\end{center}
\caption{The quasiparticle density of states with and without interactions for $W_A=2$, $W_B=1$, $v=0.2$, $\Delta=0.2$ (eV) and $n_e=1.5$ corresponding to LiV$_2$O$_4$.
The inset represents the case of LiTi$_2$O$_4$ as $n_e=0.5$.
}
\label{fig1}
\end{figure}
\end{minipage}

\begin{minipage}{8.2cm}
\begin{figure}
\begin{center}
\epsfxsize=8cm \epsfbox{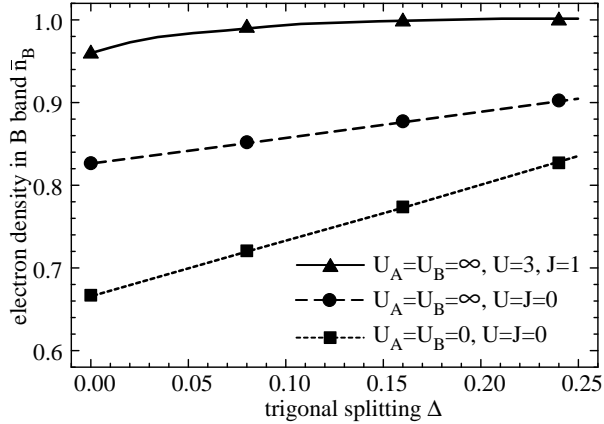}
\end{center}
\caption{The $\Delta$ dependence of electron density in $B$ band for $W_A=2$, $W_B=1$, $v=0.2$ (eV) and $n_e=1.5$.
The comparison among three cases shows that the inter-band interactions effectively works to stabilize integral valence even for rather small $\Delta$.}
\label{fig2}
\end{figure}
\end{minipage}

\begin{minipage}{8.2cm}
\begin{figure}
\begin{center}
\epsfxsize=8cm \epsfbox{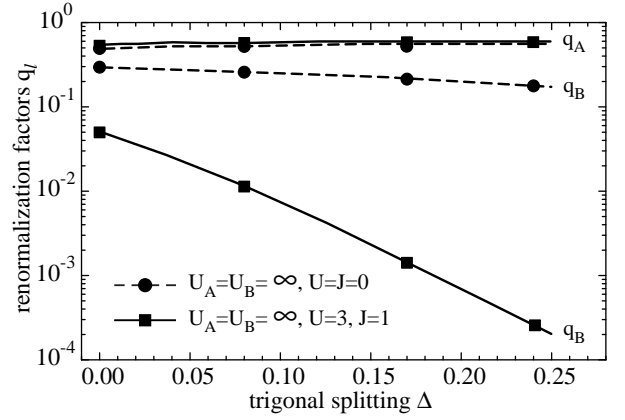}
\end{center}
\caption{The $\Delta$ dependence of renormalization factors for $W_A=2$, $W_B=1$, $v=0.2$ (eV) and $n_e=1.5$.
The $B$ band is highly renormalized by the inter-band interactions.}
\label{fig3}
\end{figure}
\end{minipage}

\begin{minipage}{8.2cm}
\begin{figure}
\begin{center}
\epsfxsize=8cm \epsfbox{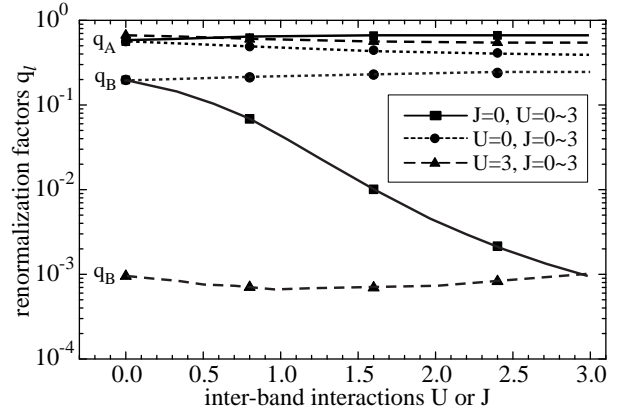}
\end{center}
\caption{The inter-band interaction dependences in the renormalization factors for $W_A=2$, $W_B=1$, $v=0.2$, $\Delta=0.2$ (eV) and $n_e=1.5$.
}
\label{fig4}
\end{figure}
\end{minipage}

\end{multicols}
\end{document}